\documentclass[12pt]{article}
\usepackage[dvips]{graphicx}
\usepackage{epsfig}
\usepackage{amsmath}
\usepackage{bm}
\usepackage{times}
\makeatletter
\@addtoreset{equation}{section}
\renewcommand\theequation{\thesection.\@arabic\c@equation}
\makeatother
\newcommand{\ens}{\epsilon_{ns}}
\newcommand{\es}{\epsilon_{s}}

\newcommand{\vfus}{\mathcal{V}^{us}_\mathit{eff}}
\newcommand{\vfub}{\mathcal{V}^{ub}_\mathit{eff}}
\newcommand{\vfud}{\mathcal{V}^{ud}_\mathit{eff}}
\newcommand{\afus}{\mathcal{A}^{us}_\mathit{eff}}
\newcommand{\afud}{\mathcal{A}^{ud}_\mathit{eff}}

\def\roughly#1{\mathrel{\raise.3ex\hbox{$#1$\kern-.75em%
\lower1ex\hbox{$\sim$}}}}
\def\lsim{\roughly<}

\begin{document}
\begin{titlepage}
\begin{center}
{\LARGE\bf  Matching Chiral Perturbation Theory 
and the Dispersive Representation of the 
 Scalar \\[2mm] 
{\boldmath$K \pi$} Form Factor}
\\[12mm]
{\normalsize\bf V\'eronique Bernard~${}^{a,}$\footnote {Email:~bernard@lpt6.u-strasbg.fr},
Emilie Passemar~${}^{b,}$\footnote{Email:~passemar@itp.unibe.ch} 
}\\[4mm]

{\small\sl ${}^{a}$ Universit\'{e} Louis Pasteur, Laboratoire de Physique Th\'{e}orique,} \\
{\small\sl 3-5 rue de l'Universit\'{e}, F-67084 Strasbourg, France}\\
{\small\sl ${}^{b}$ Institute for Theoretical Physics,
University of Bern, Sidlerstr. 5, CH-3012 Bern, Switzerland}\\
[12mm]
\end{center}

\noindent{\bf Abstract:}
\noindent
We perform a matching of the two loop-chiral perturbation theory
representation of the scalar $K \pi$ form factor to a dispersive 
one. Knowing the value of $F_K/F_\pi$ and $f_+(0)$ in the Standard Model (SM) 
allows to determine  two $O(p^6)$ 
LECs, the slope of the scalar form 
factor and  the deviation of the Callan-Treiman theorem. Going beyond the 
SM 
and assuming the knowledge of the slope of the scalar form 
factor from experiment, 
the matching  allows us to determine the ratio of $F_K/F_\pi$, $f_+(0)$,
a certain combination of non-standard couplings, the deviation of the Callan-Treiman theorem and the  two $O(p^6)$ LECs.
\end{titlepage}
\pagenumbering{arabic}
\renewcommand{\thefootnote}{\arabic{footnote}}
\parskip12pt plus 1pt minus 1pt
\topsep0pt plus 1pt
\setcounter{totalnumber}{12}
\section{Introduction}
One privileged framework for studying meson and baryon properties in the 
low-energy domain is chiral perturbation theory (ChPT), the effective field
theory of the Standard Model (SM). It is well known that it involves
so-called low energy constants (LECs) which describe the influence of ``heavy''
degrees of freedom not contained explicitly in the Lagrangian.
Determining these LECs is a difficult non-perturbative problem. 
It is, however, extremely important  to pin them down  in order to reach 
predictivity. Different attempts are
made: phenomenological evaluation based on experimental 
information at low energies, resonance saturation, 
sum rules, resonance chiral theory, lattice QCD as well as
matching \cite{be07}. Here we will be concerned with two QCD quantities,
the pion and kaon decay constants, $F_\pi$ and $F_K$ respectively and  
two of the $O(p^6)$ LECs $C_{12}$ and $C_{34}$ \cite{bce99}. These last two 
enter the calculation of two very important quantities, namely 
the strangeness changing vector and scalar form factors
in ChPT at two loops. For example, the knowledge of the 
scalar form factor at the so-called Callan-Treiman (CT) point as well as the 
one of the vector form factor at zero momentum transfer 
enable one to test the SM~\cite{Palutan07,bm06,bops06}. There are thus many theoretical
works related to the extraction of these quantities \cite{bops06}-\cite{boy07}.
Also they are extensively investigated in the four experiments by NA48~\cite{NA48e, NA48mu}, KLOE~\cite{KLOEe, KLOEmu}, 
KTEV~\cite{KTeVe}
and ISTRA~\cite{ISTRAe}. A determination of the two $O(p^6)$ LECs $C_{12}$ and $C_{34}$ has
already been done for example in Refs.~\cite{jop04,ceekpp05} using some a priori 
experimental knowledge of 
the pion and kaon decay constants. Here we want to go somewhat further.
It was realized in Ref.~\cite{bops07} that, independently of the problems
related to quark mixing, the actual values of  
these two decay 
constants are known only if one assumes the electroweak couplings of the SM. We want to
investigate some consequences of this observation.  For this, 
we will use the dispersive representation of the $K_{\mu 3}$ scalar
form factor introduced in Ref.~\cite{bops06} and do a matching to the 
two-loop calculation of Bijnens and Talavera \cite{Bijnens:2003uy}. That is
we will concentrate here on standard ChPT. 
Would the SU(3) quark condensate be much smaller than the SU(2) one  
as discussed in Refs.~\cite{mou00,dsg00,DescotesGenon:2007qi}
would the results presented here be different. 
A study of this is
beyond the scope of the letter. 
From the matching
and assuming the SM, we will be able to determine the two $O(p^6)$ 
LECs, the slope of the scalar form 
factor and  the deviation of the Callan-Treiman theorem. Going beyond the 
SM 
and assuming the knowledge of the slope of the scalar form 
factor from experiment, 
the matching will allow us to determine the ratio of $F_K/F_\pi$, $f_+(0)$,
a certain combination of non-standard couplings, the deviation of the Callan-Treiman theorem and the  two $O(p^6)$ 
LECs.

In section \ref{intro}, we discuss the decay constants and the vector
$K \pi$ form factor. We show that they are known only in the framework of the
SM and we introduce their modification from effects beyond the SM.
We write these modifications in terms of 
three parameters which describe the coupling of right-handed quarks to the W-boson as well as the modification
of the left-handed ones \cite{bops07}. We will see however that our discussion
is more general. We recall in section \ref{disp} the dispersive
representation of the scalar form factor introduced in Ref. \cite{bops06}
and in section \ref{chpt} its expression in a two-loop ChPT calculation
\cite{ps02,Bijnens:2003uy}. 
We do the matching of these two representations in section~\ref{match}
and discuss the results both in the SM and beyond in section~\ref{results}.

\section{Decay constants and vector form factor}\label{intro}

Fundamental QCD quantities are the pion and kaon 
decay constants defined as
\begin{equation}
\langle 0|A_\mu^a | M^b(p) \rangle =i \delta^{ab} F_M p_\mu \, ,
\end{equation}
with $A_\mu$ the axial current operator and $M$ the pion or the kaon mass, 
respectively.
Indeed $4 \pi F_\pi$ for example is the scale  beyond which
ChPT is not applicable anymore  and thus enters
naturally any ChPT calculations. It is common to use in these calculations
$F_\pi=92.4$~MeV and $F_K/F_\pi=1.22$. The value for $F_\pi$ ($F_K$)
comes from the (radiative) inclusive decay rates for $\pi (K) \to \mu \nu (\gamma)$
\cite{s06}. Taking their ratio leads to the value of $F_K/F_\pi$ just given. However the knowledge of
these quantities involves the axial EW couplings of quarks to the W-boson. In order to 
determine them, one thus has to know these couplings. At present the only 
well-known quantity is the vector coupling $\vfud$ of the $u$ and $d$ quarks to W. 
It is very accurately determined from $0^+ \to 0^+$ transitions in 
nuclei assuming conservation of the vector current. Its value has been very
recently updated \cite{th07} and is one standard deviation larger than in Ref. \cite{masi06} with an uncertainty one third smaller,
\begin{equation} 
\vfud =0.97418(26).
\label{eq:vfud}
\end{equation} 
($\vfud$ is also determined from the measurement of the neutron
life time or pionic decays \cite{Hard07} but with a much larger uncertainty).
Note that though the numerical results of this letter would be slightly 
affected by a small change in $\vfud$, the conclusions would not be modified.
Thus what can presently be given very precisely are the values of the pion and kaon decay constants
in the SM where the axial and vector couplings
are equal. Physics beyond the Standard Model can lead to 
a small difference between the axial and vector couplings leading
to some small contributions from right-handed currents (RHCs). Such a scenario
has been discussed in Ref. \cite{bops07} where three small parameters
$\ens, \es$ and $\delta$ enter naturally
into an effective non-quite decoupling theory beyond the leading order (LO) 
\cite{hs04}. The first 
two describe such couplings of RHCs to non-strange and  
strange quarks to W while the last one modifies the left-handed 
couplings. We refer to Refs. \cite{bops06,bops07} for a more
thorough discussion of these quantities. 
Let us just write here the modification of the vector and axial couplings
at next-to-leading order (NLO) of this effective theory: 
\begin{eqnarray}
|\vfud|^2 &=& \cos^2\hat\theta \, , \nonumber\\ 
|\afud|^2 &=& \cos^2\hat\theta\,( 1 - 4 \, \ens) \, , \nonumber\\
|\vfus|^2 &=& \sin^2\hat\theta\, \left( 1 +
 2\frac{\delta+\ens}{\sin^2\hat\theta} \right) (1 + 2\, \es - 2\, \ens) \, , \nonumber\\
|\afus|^2 &=& \sin^2\hat\theta\, \left( 1 +
 2\frac{\delta+\ens}{\sin^2\hat\theta}\right) (1 -2\, \es - 2\, \ens)~.
\label{effcouplings}
\end{eqnarray}
In these expressions and in the following, the hat on a quantity denotes that 
its value is determined from the measured semi-leptonic branching ratio
assuming the SM electroweak couplings. 
We also introduced here 
the Cabibbo angle $\hat \theta$ neglecting in the SM the $ub$ CKM matrix 
element as suggested by the measurement of $\vfub$.
With these expressions, one gets:
\begin{equation}
|{\vfud}|^2 +|{\vfus}|^2 \equiv 1+\Delta_{{\rm {unitarity}}}=1+2(\delta+\ens) +2(\es-\ens) \sin^2 \hat
\theta,
\label{unitarity}
\end{equation}
that is a small deviation from unitarity can occur for the vector effective 
couplings of the effective theory. 
Using the relations above one obtains for the pion and kaon decay constants 
\begin{eqnarray}
|F_{\pi}|^2 &=& \hat{F}^2_{\pi}~(1 + 4\, \ens)\nonumber \\
\left(\frac{F_{K}}{F_{\pi}}\right)^2 
&= &\left(\frac{\hat{F}_{K}}{\hat{F}_{\pi}}\right)^2 
\frac{\sin^2\hat\theta}{\cos^2\hat\theta} \frac{|\afud|^2}{|\afus|^2}
 =\left (\frac{\hat{F}_{K}}{\hat{F}_{\pi}}
\right)^2 \frac{1+2\,(\es-\ens)}{1+\frac{2}{\sin^2\hat{\theta}}
(\delta+\ens)}~,
\label{FKFpi}
\end{eqnarray}
where 
\begin{equation}
\hat{F}_{\pi}=\left( 92.3\pm 0.1 \right) \, {\rm MeV}~,\,\,\,\,\, \,\,\,\,\, \hat{F}_K/\hat{F}_\pi= 1.192 \pm 0.007 \, .
\label{hatFKFpi}
\end{equation}
The value of $\hat{F}_K/\hat{F}_\pi$ is thus markedly smaller than what has been used so far in ChPT. It 
is  obtained from 
the ratio $\Gamma_{K^+_{l2}(\gamma)}/ \Gamma_{\pi^+_{l2}(\gamma)} = 1.3383 (46) $ \cite{bm06} of the inclusive decay rates for $\pi (K) \to \mu \nu$
and the value of $\vfud$ given in Eq.~(\ref{eq:vfud}). The value of $\hat{F}_\pi$ is obtained from Refs.~\cite{ms93,am02,dm05}
\begin{equation}
\sqrt 2\hat{F}_{\pi}=\left( 130.766~\left(\frac{0.9750}{\vfud}\right) +0.156~C_1 \right) \, {\rm MeV} \, ,
\end{equation}
with $C_1=-2.56 \pm 0.5$ \cite{dm05}.

Same discussion holds for the vector form factor. 
Its knowledge at
zero momentum transfer is crucial for the determination of the 
CKM matrix element $\vfus$. One has 
\begin{equation}
|f^{K^0\pi^-}_+(0)|^2 = |\hat{f}^{K^0\pi^-}_+(0)|^2 \frac{\sin^2 \hat{\theta}}{|\vfus|^2} =
\left[ \hat{f}^{K^0\pi^-}_+(0)\right]^2 \,
\frac{1-2(\es-\ens)}{1+\frac{2}{\sin^2\hat{\theta}}
(\delta+\ens)}~,
\label{fpz}
\end{equation}
where the value obtained in the SM
\begin{equation}
\hat{f}^{K^0\pi^-}_+(0)= 0.9574(52)
\label{hatfpz}
\end{equation}
 comes from an average value of the $K_{Le3}$ 
and $K_{Se3}$ decay rate \cite{Palutan07} leading to $|f_+(0) \vfus| = 0.21615(55)$.
Note that the same denominator enters both $F_K/F_\pi$ and 
$f^{K^0\pi^-}_+(0)$ so that their ratio depends only on the difference
$\es - \ens$. Also combining Eqs.~(\ref{unitarity}) and (\ref{fpz}), one sees that at NLO of the effective theory, 
the deviation from 
unitarity of the vector couplings can be related to the difference between the
physical value of $ f^{K^0\pi^-}_+(0)$
and its hat value. One has
\begin{equation}
\Delta_{{\rm {unitarity}}}= \sin^2\hat{\theta} \left( \frac{|\hat{f}^{K^0\pi^-}_+(0)|^2}{ |f^{K^0\pi^-}_+(0)|^2} -1 \right) \, .
\label{unitfp}
\end{equation}
Clearly this deviation can only be very small, its sign depending on the 
exact value of $f^{K^0\pi^-}_+(0)$. In fact, from the lattice results, one 
expects
$-2.5 \times 10^{-3} <\Delta_{{\rm {unitarity}}} < 8 \times 10^{-4}$.

It was discussed in Ref.~\cite{bops07} that the parameters
$\ens$ and $\delta$ should be small, less than a percent. 
Note however that in Eqs.~(\ref{FKFpi}) and (\ref{fpz}) the quantity 
$\delta +\ens$ is multiplied by the not so small quantity $1/\sin^2 \hat 
\theta$,
we will thus refrain in the following from expanding the denominator in these
expressions.  
On the other
hand, $\es$ could be enhanced to a few percent level which could
be explained for example by an inverted hierarchy in right-handed 
flavour mixing. One expects from these estimates that $F_K/F_\pi$ and
$f_+(0)$ should be more affected than $F_\pi$ by the presence of RHCs.

Our discussion will in fact be more general. Indeed, in the following, we will
not consider any modification of $F_\pi$ from its value obtained
with the effective couplings of the SM. As just said these are expected to
be rather small. Thus only two quantities will play a role in the following
which can be chosen as
\begin{equation}
\alpha = \frac{1+2(\es-\ens)}{1+\frac{2}{\sin^2\hat{\theta}}
 (\delta+\ens)}\,\, \, \, \,~~~\mathrm{and}~~ \, \, \, \, \, 
\beta =\frac{1-2(\es-\ens)}{1+\frac{2}{\sin^2\hat{\theta}}(\delta+\ens)}~.
\end{equation}
They just parametrize our ignorance of the physical values of $F_K/F_\pi$ and 
$f_+(0)$ if there is physics beyond the SM. For the reader who prefers
to think in terms of these quantities it is easy to rewrite $\es-\ens$ and
$\delta + \ens$ as a function of $\alpha$ and $\beta$.

\section{Matching} 

\subsection{Dispersive representation} \label{disp} 
A dispersive representation of the scalar form factor 
was introduced in Ref.~\cite{bops06}. It is based on a twice subtracted 
dispersion relation and reads:  
\begin{equation}
\bar {f}_0(t) \equiv \frac{f_{0}^{K^0\pi^{-}}(t)}{f_{0}^{K^0\pi^{-}}(0)}=
\exp\Bigl{[}\frac{t}{\Delta_{K\pi}}(\mathrm{ln}C- G(t))\Bigr{]}~, 
\label{fDisp}
\end{equation}
with
\begin{equation} 
G(t)=\frac{\Delta_{K\pi}(\Delta_{K\pi}-t)}{\pi}\nonumber\ \int_{(M_K+M_\pi)^2}^{\infty}
\frac{ds}{s}
\frac{\phi(s)}
{(s-\Delta_{K\pi})(s-t-i\epsilon)}~.
\label{Dispf}
\end{equation}
and $\phi(s)$ the phase of the form factor. 
It has many advantages.
First, it introduces the value of the form factor at
the Callan-Treiman point $\Delta_{K \pi}=M_K^2-M_\pi^2$, a quantity $C$ 
which is not affected by chiral corrections beyond $SU(2) \times SU(2)$.
Thus these are 
of order ${\cal O}(m_u,m_d)$ while the slopes have
larger corrections of the order of ${\cal O}(m_s)$. 
Second, it allows to test the Standard Model. Indeed one
can relate the scalar form factor at the Callan-Treiman point
to the quantity $\es-\ens$.
One has:
\begin{equation}
C\equiv \bar {f}_0(\Delta_{K\pi})=\frac{F_{K}}{F_{\pi}}\frac{1}{f_{+}^{K^0\pi^-}(0)}+  
\Delta_{CT}~,
\label{C}
\end{equation}
which using Eqs.~(\ref{FKFpi}) and (\ref{fpz}), leads to
\begin{equation}
C=\frac{\hat{F}_{K}}{\hat{F}_{\pi}}\frac{1}{\hat{f}_{+}^{K^0\pi^-}(0)} (1+2(\es-\ens))+  
\Delta_{CT} = B_{exp} (1+2(\es-\ens))+ \Delta_{CT} ~.
\label{eq:chat}
\end{equation}
Hence one obtains from the values, Eqs.~(\ref{hatFKFpi}) and (\ref{hatfpz}),
\begin{equation}
\ln C= 0.2188 \pm 0.0035 + \Delta {\epsilon}
\label{eq:lnce}
\end{equation}
where $ \Delta {\epsilon} \equiv \Delta_{CT}/B_{exp} + 2(\es -\ens)  $
and $B_{exp} =1.2446 \pm 0.0041$.
Expanding $\bar {f}_{0}(t)$  
\begin{equation}
\bar {f}_{0}(t)=
1+\lambda_{0}\frac{t}{M_{\pi}^2}+\frac{1}{2}\lambda'_{0}\Bigl{(}\frac{t}{M_{\pi}^2}\Bigr{)}^2 + \cdots~,
\label{f0}
\end{equation}
the linear slope is given in terms of $\ln C$ as
\begin{eqnarray}
\lambda_{0} = \frac{M_{\pi}^2}{\Delta_{K \pi}}(\mathrm {ln C} - G(0) )~,
\label{eq:lambd0}
\end{eqnarray}
with $G(0)=0.0398 \pm 0.0036 \pm 0.0020$ \cite{bops06} whereas the curvature reads
\begin{eqnarray}
\lambda'_{0} = \lambda_{0}^2  - 2 \frac{M_{\pi}^4}{\Delta_{K\pi}} G'(0) = \lambda_{0}^2  + (4.16 \pm 0.50)\times 10^{-4}.
\label{curvature}
\end{eqnarray}
Note that in order to get a very precise description of $\bar{f}_0(t)$ over the
entire physical region it is necessary to do an expansion up to third order
\cite{bops07a}. Here we will concentrate on the region around $t=0$. 

\subsection{ChPT to two loops} \label{chpt}

The scalar form factor 
was calculated to two loops  in ChPT in Ref.~\cite{Bijnens:2003uy}.
These authors introduced 
the quantity
\begin{equation}
\label{deftildef0}
\tilde f_0(t) = f_+(t)+\frac{t}{M_K^2-M_\pi^2}
\left(f_-(t)+1-F_K/F_\pi\right)
= f_0(t)+\frac{t}{M_K^2-M_\pi^2}\left(1-F_K/F_\pi\right)
\,.
\end{equation}
The main advantage is that this quantity
has no dependence on the $L_i^r$ at order $p^4$, only via
order $p^6$ contributions. It, however, depends on the 
${\cal O}(p^6)$ LECs $C_i^r$ in the following way:
\begin{eqnarray}
\label{resultfp0}
\tilde f_0(t) &=& 1-\frac{8}{F_\pi^4}\left(C_{12}^r+C_{34}^r\right)
\left(M_K^2-M_\pi^2\right)^2
+8\frac{t}{F_\pi^4}\left(2C_{12}^r+C_{34}^r\right)\left(M_K^2+M_\pi^2\right)
\nonumber\\&&
-\frac{8}{F_\pi^4} t^2 C_{12}^r
+\overline\Delta(t)+\Delta(0)\,.
\end{eqnarray}
The quantities ${\overline{\Delta}}(t)$ and
$\Delta(0)$ have contributions from loops, thus depend on $F_\pi$,
 and from the 
LECs $L_i$. Note that $L_5$ is related to $F_K/F_\pi$.  $\overline\Delta(t)$ and
$\Delta(0)$ 
can in principle be calculated
to order $p^6$ accuracy with the knowledge of the $L_i^r$ to order $p^4$
accuracy. $\overline\Delta(t)$ has been parametrized in the physical region
as:
\begin{eqnarray}
\overline\Delta(t) &=& -0.25763 t  + 0.833045 t^2 +  1.25252 t^3
\quad [K^0_{e3}],
\nonumber\\
\overline\Delta(t) &=&  -0.260444 t  + 0.846124 t^2 + 1.33025 t^3
\quad [K^+_{e3}].
\label{deltat}
\end{eqnarray}
Different sets of $L_i^r$ have been obtained from a fit to $K_{\ell4}$ data
to two loops \cite{bijyy}. The error from the values of the different sets of $L_i^r$
is about 0.0013 at $t=0.13$~GeV$^2$.
Contributions from the loops and the $L_i^r$ to $\Delta(0)$ are:
\begin{equation}
\Delta(0) = -0.0080\pm0.0057[\mbox{loops}]\pm0.0028[L_i^r]\,,
\label{Deltaz}
\end{equation}
where  the central value arises from a cancellation between ${\cal{O}}(p^4)$
and ${\cal{O}}(p^6)$ terms $-0.008=-0.02266\, (p^4) \, +0.01130 \, 
(p^6 \, {\rm { pure \, loops}}) +0.00332 \, (p^6 \, L_i)$.
For more details, see Ref. \cite{Bijnens:2003uy}.

\subsection{Basic Formulae} \label{match}

Relating the dispersive representation to the two-loop ChPT
calculation will allow us to determine the deviation from the
Callan-Treiman theorem, $F_K/F_\pi$, the LECs $C_{12}$ and $C_{34}$  
as well as either the slope of the form factor or the quantity $\delta +\ens$ once one has fixed the quantities $\es
-\ens$ and either $\delta +\ens$ or the slope of the form factor, respectively.
Taking the derivative of
Eq.~(\ref{resultfp0}), the ChPT expression for the slope is:
\begin{equation}
\lambda_0 f_+(0)= \frac{M_\pi^2}{\Delta_{K\pi}} \left( \frac{F_K}{F_\pi} -1 \right) + \frac{8M_\pi^2 \Sigma_{K\pi}}{F_\pi^4}(2C_{12}+C_{34})+M_\pi^2\overline{\Delta}'(0)  \,\,\, ,
\label{eq:lamb0}
\end{equation}
with $\Sigma_{K \pi}=M_K^2+M_\pi^2$.
 Combining the curvature obtained from
Eq.~(\ref{resultfp0}),
\begin{equation}
\lambda_0'f_+(0)=-\frac{16M_\pi^4}{F_\pi^4}C_{12}+M_\pi^4 \overline{\Delta}''(0)~,
\end{equation}
with
the two-loop result for $f_+(0)$ 
\begin{equation}
f_+(0)= 1+ \Delta(0) - \frac{8}{F_\pi^4} (C_{12}+ C_{34}) \Delta_{K \pi}^2~,
\end{equation}
one gets an expression for $2 C_{12} + C_{34}$. Inserting it into Eq.~(\ref{eq:lamb0}),
using further the dispersive relation, Eq.~(\ref{curvature}) and expressing $f_+(0)$ and 
$F_K/F_\pi$ in terms of the hat quantities, Eqs.~(\ref{FKFpi}) and (\ref{fpz}), one obtains 
a second order equation for the slope $\lambda_0$ whose solution reads:
\begin{equation}
\lambda_0 = -\frac{M_\pi^2}{\Sigma_{K \pi}} 
\left(1 - \sqrt {1 - 2 \frac{\Sigma_{K \pi}^2}{\Delta_{K \pi}} \left( \frac 
{Y}{\Delta_{K \pi}} - G'(0) \right) }   \right)
\label{eq:lambda0}
\end{equation}  
with 
\begin{eqnarray}
 & &  Y = 1  -
\frac{\Delta_{K \pi}}{\Sigma_{K \pi}} \frac{\hat{F}_K}{\hat{F}_{\pi} \hat{f}_+(0)}
(1 + 2 (\es -\ens)) 
\\ \nonumber
& &- \frac{1}{\hat{f}_+(0)} 
\left( 1+ \Delta (0) +
\frac{\Delta_{K \pi}^2}{2} \overline \Delta ''(0)- \frac{\Delta_{K \pi}}{\Sigma_{
K \pi}} \left(1-\Delta_{K \pi} \overline \Delta '(0)\right)  \right)(1+\es-\ens)
\sqrt{1+y} ~.
\label{eq:bigy}
\end{eqnarray} 
Contrary to $\ln C$ which depends only on $\es -\ens$, $\lambda_0$ is a function of 
both quantities $ \es -\ens$ and $ y=2
(\delta+\ens)/\sin^2\hat{\theta}$. 
Once $\lambda_0$ is known, all the other quantities are determined in terms
of $\es-\ens$ and $y$. $F_K/F_{\pi}$, $f_+(0)$ are given by Eqs.~(\ref{FKFpi}) and (\ref{fpz}) respectively and
\begin{eqnarray}\label{c1234}
C_{12} &=&\frac{F_\pi^4}{16} \left( - \frac{\lambda_0' f_+(0)}{ M_\pi^4}
+ \overline \Delta ''(0) \right)~,
\\ \nonumber
C_{34} &=&\frac{F_\pi^4}{8 \Delta_{K\pi}^2} \left( 1 +\Delta(0) -f_+(0) \right) -
C_{12} \, .
\\ \nonumber
\end{eqnarray}
One has trivially from Eqs.~(\ref{C}) and (\ref{eq:lambd0})
\begin{equation}
\Delta_{CT}= B_{exp}\left( \frac{\Delta_{K \pi}}{M_\pi^2} \lambda_0 +G(0) -\ln B_{exp} -2(\es
-\ens) \right) \, .
\end{equation}
\section{Results and Conclusion} \label{results}

We will not try  here to get exact results but more trends of what 
can be expected from such a matching. Indeed, in order to do the 
matching, one has to use values for $\Delta(0)$ and $\overline {\Delta} (t)$
which have been determined using $F_\pi=92.4$ MeV and $F_K/F_\pi=1.22$.
Thus our results will not be completely consistent since we will in the 
following
determine $F_K/F_\pi$ from Eq.~(\ref{FKFpi}). Also if $\ens \neq 0$, $F_\pi$
will be modified, see Eq.~(\ref{FKFpi}). However,  
we do not expect much changes in the result would one do a consistent 
calculation. Indeed in  $\Delta(0)$ the contribution from the 
$L_i$ is rather small and a small uncertainty was found in $\overline {\Delta} (t)$
while using different sets of $L_i$'s, see also Ref. \cite{bij07a}. Besides, as 
already mentioned 
one  expects values of $\ens$ smaller than a percent so that $F_\pi^2$ would
be changed by at most $4 \%$. All these effects can, to our opinion, very well 
be accounted by
the rather conservative uncertainties given for $\Delta(0)$, Eq.~(\ref{Deltaz}). 
We will thus vary $\Delta(0)$ within its error bars
to see how the results are affected. 
Ultimately, we would of course like to study the dependence of the results 
on $F_\pi$ since it would
enable one to determine independently $\delta$ and $\ens$. 
It would indeed be very interesting to test the  quark-lepton
universality
which implies  $\delta=0$ \cite{st07}. 
However the conservative uncertainty on $\Delta(0)$, Eq.~(\ref{Deltaz}), is too big, as we will see, to really get 
very precise results. Note also that since the 
fits were done in Ref. \cite{bijyy}, new $K_{\ell4}$ data are available. New fits 
should certainly be performed~\cite{bij07a} leading to an updated value for $\Delta(0)$.


\begin{table}[t]
 \begin{center}
    \begin{tabular}{|c|cc|cccccc|}
      \hline 
$\Delta(0)$&$\es-\ens$ &$\Delta_{{\rm{unitarity}}}$& $\lambda_0$ & $\Delta_{CT}$ & $f_+(0)$ & $F_K/F_\pi$ &  $C_{12} $ & $C_{34}$  \\
& &$(10^{-3})$&$(10^{-3})$&$(10^{-2})$&&&$(10^{-6})$&$(10^{-6})$ \\
\hline
-0.008&SM &SM& $15.20$ & $-0.118$ &\,\,\, 0.957 * & \,\,\,\, 1.192 * &  $-0.421$&$6.480$ \\
&$0$ & $\!\!\!\!-1.5$& $15.03$& $-0.368$ & 0.972  & $1.210$&  $-0.484$ &$ 3.971$\\

&$0$ & $\!\!\!\!-3.1$& $14.85$& $-0.622$ & 0.987  & $1.229$&  $-0.550$ &$ 1.344$ \\

&$0$ & $1.5$& $15.37$& $\,\,\,0.127$ & 0.943  & $1.174$&  $-0.362$ &$ 8.879$ \\
&$0$ & $3.1$& $15.53$& $\,\,\,0.369$ & 0.930  & $1.157$&  $-0.306$ &$\! 11.176$ \\
 \hline
-0.0165&SM &SM& $14.46$ & $-1.193$ &\,\,\, 0.957 * &\,\,\,\, 1.192 * & $-0.170$&$4.741$ \\
&$0$ & $\!\!\!\!-1.5$& $14.30$& $-1.428$ & 0.972  & $1.210$&  $-0.235$ &$ 2.235$\\
      \hline 
\hline
0.0005&SM &SM& $15.93$ & $\,\,\,0.948$ &\,\,\, 0.957 * & \,\,\,\,1.192 *&  $-0.683$&$8.229$ \\
&$0$ & $\!\!\!\!-1.5$& $15.75$& $\,\,\,0.684$ & 0.972  & $1.210$&  $-0.743$ &$ 5.718$\\
      \hline 

    \end{tabular}
    \caption{Values of the slope of the form factor $\lambda_0$, the deviation from the Callan-Treiman theorem 
$\Delta_{CT}$, the value of the vector form factor at zero momentum
transfer $f_+(0)$,
the ratio of the pion and kaon decay
constants $F_K/F_\pi$, two ${\cal{O}}(p^6)$ LECs $C_{12}$ and $C_{34}$
as a function of the non standard couplings $\es -\ens$ to the
$W$-boson and the deviation from unitarity $\Delta_{{\rm {unitarity}}}$ of the effective couplings. 
The star means that the quantities are known from experiment, Eqs.~(\ref{hatFKFpi}) and (\ref{hatfpz}).
The dependence on the ChPT input quantity $\Delta(0)$
is also shown.}  
    \label{tab:Kl3formfactors}
  \end{center}
\end{table}

In the following, we will be using the central value for
$\Delta(0) = -0.008$, for $\overline \Delta (t)$ the values
from the fit to  neutral kaons and $ 2 M_\pi^4 G'(0)/\Delta_{K \pi}=-4.66  \times 10 ^{-4}$. We will also consider the deviation from unitarity of 
the vector effective couplings, $\Delta_{{\rm{unitarity}}}$, Eq.~(\ref{unitarity}) instead of the quantity $\delta +\ens$. It is easy to recover the 
values of this quantity from Eq.~(\ref{unitarity}) if needed.
We will consider two different scenarios. In the first one, we will fix $\es -\ens=0$ and study the dependence of the results on $\Delta_{{\rm{unitarity}}}$. In the second one, we will study the case  $\es - \ens \neq 0$.

$\bullet$ First, we will 
assume that we are in the SM. In that case, $\delta=\ens=\es=0$.
The results are given in table \ref{tab:Kl3formfactors}.
$F_K/F_\pi$ and $f_+(0)$ are the hat quantities determined from experiments
as discussed in section \ref{intro}, see Eqs.~(\ref{hatFKFpi}) and (\ref{hatfpz}). With
the updated value of $\vfud$, they are now in good agreement with the 
recent lattice results for
$F_K/F_\pi=1.189(7)$ \cite{milc07} and 
$f_+(0)=0.9609(51)$ \cite{ukqcd07} obtained 
with staggered and DWF
fermions respectively. 
Note however that the value of $F_K/F_\pi$ from Ref. \cite{milc07}
is somewhat on the lower side of most of the lattice results. 
A rather small value for $F_K/F_\pi$ has been obtained recently from the CP-PACS/JLQCD
collaboration, however most of the SU(3) lattice results give 
central values around $1.21$, see Refs.~\cite{kan07,bloi07}.  
Lattice values for  $f_+(0)$ are $0.95 < f_+(0)<0.98$ \cite{kan07,mes07}
while the widely used quark model of Leutwyler and Roos \cite{lero84} gives
$f_+(0)=0.961 \pm 0.008$. $\lambda_0$  is on the large side of the
experimental results while consistent with the KLOE result as obtained
from a linear parametrization for the scalar form factor and a quadratic one for the vector \cite{KLOEmu}. It has however recently been understood that
the use of a linear parametrization is not appropriate. It leads to  
a value for the slope of the scalar form factor  larger than it actually is
\cite{franz07}.
$\Delta_{CT}$ is 
very small
as expected from the NLO result in ChPT in the isospin limit~\cite{gale85}
\begin{equation}
\Delta_{CT}^{NLO}= (-3.5 \, \pm 8) \cdot 10^{-3}
\label{DeltaCTNLO}
\end{equation} 
where the error is a conservative estimate assuming some typical 
corrections of ${\cal O}(m_{u,d})$ and ${\cal O}(m_s)$ \cite{leutpriv}.
The LEC $C_{12}$ is found to be negative. 
 Resonance
exchange models give negative values of the order
of $10^{-5}$ for a scalar mass exchange of $M_S \sim 980$ MeV which corresponds to the $a_0$. Other masses have also been considered \cite{ceekpp05}.
Taking $M_S$ between 1 GeV and 1.5 GeV one gets $-9 \cdot 10^{-6}
\lsim C_{12} \lsim -1.8 \cdot 10^{-6}$ . Assuming that
the  LECs determined within
these resonance exchange models correspond to  a scale equal to $M_S$
and evolving them to the $\rho$ scale one gets values between $-7.8
\cdot 10^{-6}$ and $4.0 \cdot 10^{-6}$ for the range of the scalar masses
discussed above \cite{jop04}. In that reference, $C_{12}=(0.3 \pm 5.4) \cdot
10^{-7}$ for a value of $\lambda_0 =0.0157\pm0.0010$ where the central
value corresponds to $f_+(0)=0.976$. This is consistent
with our findings within the error bars. However they have a smaller result
for the sum $(C_{12}+C_{34})(M_\rho)=(3.2 \pm 1.5) \cdot 10^{-6}$. Thus
calculating the $C_i$'s contribution
to $f_+(0)$
\begin{equation}
f_+(0)=-\frac{8}{F_{\pi}^4} ( C_{12} +C_{34}) (M_K^2-M_\pi^2)^2 \, ,
\label{eq:fp0ci}
\end{equation}
our result is twice as large in absolute value than the one given 
in that letter or in the pioneering work 
\cite{lero84},
$f_+(0)=-0.016 \pm 0.008$.
In the case of $\Delta_{CT}$, the $C_i$'s contribution is given by:
\begin{equation}
\Delta_{CT}|_{C_i}=\frac{16}{F_{\pi}^4} (2 C_{12} +C_{34}) M_\pi^2 (M_K^2-M_\pi^2) \, .
\label{eq:deltactci}
\end{equation}
Subtracting it to the value of $\Delta_{CT}$ given in the table, one finds $\Delta_{CT}-
\Delta_{CT}|_{C_i}=-6.68 \cdot 10^{-3}$ in very good agreement with the two loop contribution 
recently evaluated in Ref. \cite{bij07b}, as it should. Note that adding to the 
expansion, Eq.~(\ref{f0}), the $t^3$ term from Eq.~(3.10), one obtains a good 
parametrization of Eq.~(\ref{fDisp}) up to the Callan-Treiman point.

$\bullet$
Giving a small value to $\delta+\ens$ while keeping $\es-\ens=0$,
that is breaking the unitarity of the vector couplings, Eq.~(\ref{unitarity})
by a small amount,
the value for $\lambda_0$ given in the second entry in table 
\ref{tab:Kl3formfactors} is consistent with the one obtained in 
Ref. \cite{jop06} and  calculated along the line of a dispersion theoretical 
approach of Ref. \cite{jop00}. In this framework where, differently
from the one discussed here, a two channel approach has been used and
only one subtraction is performed, one needs two external input parameters.
These authors use the value of the form factor at zero momentum and its
 value at the CT point. With  $f_+(0) =0.972 (12)$ and $F_K/F_\pi=1.203 (16)$,
they get $\lambda_0=0.0147(4)$. Varying
$f_+(0)$ within the bounds from the lattice results one obtains
$0.0148 \lsim \lambda_0 \lsim 0.0154$. As one decreases $\Delta_{{\rm {unitarity}}}$,
the values of $\lambda_0$, $ \Delta_{CT}$ and the two ${\cal{O}}(p^6)$ LECs $C_{12}$ and 
$C_{34}$ decrease while the ones of $f_+(0)$ and $F_K/F_\pi$ increase. One 
observes a strong dependence of $\Delta_{CT}$ and $C_{34}$ on  $\Delta_{{\rm {unitarity}}}$ 
or equivalently, see Eq.~(\ref{unitfp}), on $f_+(0)$. 
 In the expression of $C_{12}$, Eq.~(\ref{c1234}), the first term on the right-hand side is negative 
and the second is positive. It turns out that both terms are of the same order 
of magnitude so that the sign and the value of $C_{12}$ result from a delicate
balance between the two terms. Here we have kept $\overline \Delta''(0)$ 
fixed from the fit to neutral kaons, first line
Eq.~(\ref{deltat}), so that the different values obtained in the tables for $C_{12}$
are only due to the changes in $\lambda_0$ and $f_+(0)$. Using for 
$\overline \Delta''(0)$ the value obtained in the fit to charged kaons
would lead to a small change in  the results. For example with this value 
one gets
 $C_{12}=-0.322 \cdot
10^{-6}$ in the SM, first line in table \ref{tab:Kl3formfactors}.
Concerning $C_{34}$ the first term is very sensitive
to the difference between $1+\Delta(0)$ and $f_+(0)$ which leads to the large
observed variations in its values.
\begin{table}[t]
 \begin{center}
    \begin{tabular}{|c|cc|cccccc|}
      \hline 
$\Delta(0)$&$\es-\ens$ &$\lambda_0$& $\Delta_{{\rm{unitarity}}}$ & $\Delta_{CT}$ & $f_+(0)$ & $F_K/F_\pi$ &  $C_{12} $ & $C_{34}$  \\
& &$(10^{-3})$&$(10^{-3})$&$(10^{-2})$&&&$(10^{-6})$&$(10^{-6})$ \\
\hline
-0.008&$-0.005$ & $14.00$ & $-2.804$ &$-0.623$ & $0.984$ & $1.213$ &  $-0.234$ & $\,\,\,\,1.534$ \\
&$-0.032$& $\,\,\, 9.01$& $-3.148$& $-1.178$& $0.987$& $1.152$& $\,\,\,\,\, 1.107$ &$-0.216$ \\
      \hline
 \hline
-0.0165&$-0.0012$ & $13.99$ & $-2.416$ &$-1.579$ & $0.980$ & $1.218$ &  $-0.202$ & $\,\,\,\,0.666$ \\
&$-0.028$& $\,9.00$& $-2.760$& $-2.130$& $0.983$ & $1.157$& $\,\,\,\, 1.132$ &$ -1.092$ \\
      \hline
\hline
0.0005&$-0.0088$ & $14.00$ & $-3.191$ &$\,\,\,\,0.325$ & $0.988$ & $1.209$ &  $-0.264$ & $\,\,\,\,2.400$ \\
&$-0.0358$& $\,\,\,9.01$& $-3.535$& $-0.234$& $0.991$ & $1.148$& $ \,\,\,\,1.084$ &$ \,\,\,\,0.659$ \\
      \hline 

    \end{tabular}
    \caption{Values of the deviation from unitarity $\Delta_{{\rm {unitarity}}}$, the deviation from the Callan-Treiman theorem 
$\Delta_{CT}$, the value of the vector form factor at zero momentum
transfer $f_+(0)$,
the ratio of the pion and kaon decay
constants $F_K/F_\pi$, two ${\cal{O}}(p^6)$ LECs $C_{12}$ and $C_{34}$ 
as a function of $\es -\ens$ and the slope of the form factor where $\es -\ens$ 
is fixed from the measurement of $\Delta \epsilon$ as explained in the text.
The dependence on the ChPT input quantity $\Delta(0)$
is also shown.}  
    \label{tab:Kl3lformfactors}
  \end{center}
\end{table}

\begin{figure}[h!]
\begin{center}
\includegraphics*[scale=0.4,angle=-90]{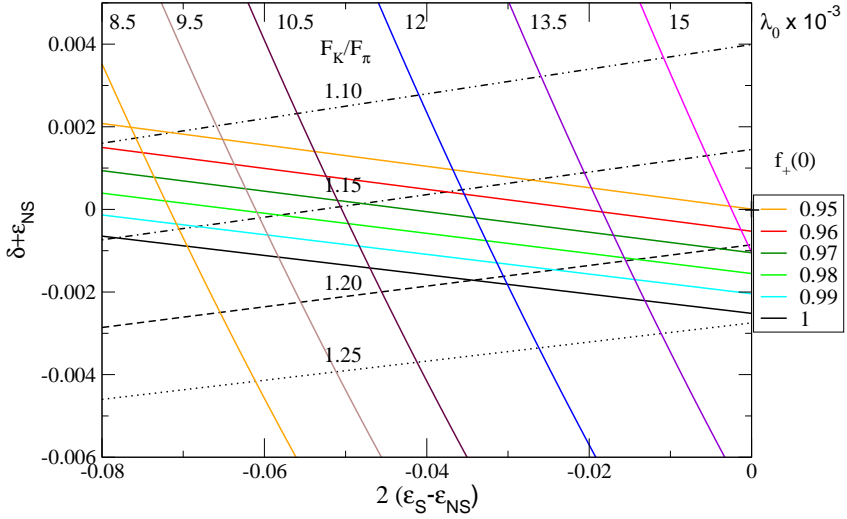}
\caption{Lines of constant values for $F_K/F_\pi$, $f_+(0)$ as in Ref.~\cite{bops07}
and $\lambda_0$ in the plane $\delta+\ens$ and $2(\es-\ens)$. $\lambda_0$ 
is calculated with the central value of $\Delta(0)$. Error on this 
quantity is larger than the one on $F_K/F_\pi$ and $f_+(0)$, see discussion
in the text.}\label{fig:RHC}
\end{center}
\end{figure}

$\bullet$ In order to get smaller values of $\lambda_0$ as demanded
by the central values of the NA48 and KTEV experiments as well as the KLOE one~\cite{KLOEmu}
when analyzed with the dispersive representation discussed in section 
\ref{disp},
one must allow for $\es -\ens \ne 0$.  
Let us first assume the NA48 result~\cite{NA48mu} which is 5 $\sigma$ deviation away from the SM one. 
The strategy
here will be  to reproduce the measured
slope $\lambda_0=(8.88 \pm 1.24) \times 10^{-3}$ 
from the dispersive analysis as well as the measured deviation from the 
Callan-Treiman theorem $\Delta \epsilon=-0.075 \pm 0.014$, Eq.~(\ref{eq:lnce}). 
This leads to a negative value of  $\es -\ens$  of the order of a few percent 
while $\delta + \ens$ has to be extremely small 
and positive.
As illustration, we show the results for $\lambda_0=9.0 \times 10^{-3}$ in table 
\ref{tab:Kl3lformfactors}. 
This leads to values for $F_K/F_\pi$ and $f_+(0)$ respectively, on the lower side of, somewhat larger than the lattice results. 
$f_+(0)$ is now much larger than in Ref. \cite{lero84} but in agreement with 
Ref. \cite{ceekpp05}.
$\Delta_{CT}$ turns out to be larger in absolute value 
than the NLO ChPT result, Eq.~(\ref{DeltaCTNLO}), however, it is within  the expected uncertainty
from higher orders. It leads to 
$\Delta \epsilon =-0.073$. Interestingly the LEC $C_{12}$ is now
much larger and positive. 
On the contrary, $C_{34}$ becomes much smaller as one goes from the
standard case to the NA48 result. Subtracting again the $C_i$'s contribution,
Eq.~(\ref{eq:deltactci}), 
to $\Delta_{CT}$ one now obtains a value twice as large as the quoted
two loop results of Ref. \cite{bij07b} due to the smaller value of $F_K/F_\pi$. 
In the first entry of table \ref{tab:Kl3lformfactors}, we give the result
corresponding to the recent determination of the slope of the form factor
by KLOE \cite{KLOEmu} using the dispersive parametrization. One can
easily calculate what is their  experimental value of $\Delta \epsilon$,
using Eqs.~(\ref{eq:lambd0}) and (\ref{eq:lnce}). 
This leads to $\Delta \epsilon=-0.015 \pm 0.025$. The $C_i$'s
contribution to $f_+(0)$ and $\Delta_{CT}$ is respectively $-0.0074$ and 
$0.0010$.

In both tables, we give results for larger and smaller
values of $\Delta(0)$, corresponding to what is the dominant uncertainty in Eq.~(\ref{eq:lambda0}). 
For comparison, in table {\ref{tab:Kl3formfactors}, we use the
same values of $\es-\ens$ and $\delta+\ens$ in all cases
so that $F_K/F_\pi$ and $f_+(0)$ are the same when varying $\Delta(0)$. 
The change in its value leads to a rather large shift
in $\lambda_0$, $\Delta_{CT}$, $C_{12}$ 
and $C_{34}$.
Thus the
conservative uncertainty on the value of $\Delta(0)$ is unfortunately too big to really enable 
one to pin down these quantities with a very good precision.
As can be seen, the matching together with  all the experimental results
on the slope of the scalar form factor available today fix the sign 
of $\es -\ens$ to be negative. With the effective couplings of the SM, 
$\lambda_0$ varies between $14.3 \times 10^{-3}$ and $16.0 \times 10^{-3}$, that is
the dependence with $\Delta(0)$ is large but can never afford such a small 
value as reported by the NA48 experiment. In table \ref{tab:Kl3lformfactors}, 
we choose to keep $\lambda_0$ and $\Delta \epsilon$ approximately fixed. The NA48 and
KLOE results from the dispersive analysis lead to values for $f_+(0) \sim  0.98$
in agreement with Ref. \cite{ceekpp05} while $F_K/F_\pi$ is rather small in the NA48 case. 
Let us mention
here that with such a small value of $F_K/F_\pi$ the value
of $\Delta(0)$ to be used should be closer to $-0.0165$ than to
$-0.008$. Indeed the contribution of $L_5$ to $f_+(0)$ is positive
\cite{bijsite}. One has in the case of the neutral kaons
\begin{equation}
f_+(0)=f_+(0)|_{{\rm{without}} L_5} -0.4136 L_5 + 5715.11 L_5^2 \, ,
\end{equation}
where the coefficient of $L_5^2$ is $-8 \left(M_K^2-M_\pi^2 \right)^2/F_\pi^4$, i.e. the same 
as the one of $C_{12}+C_{34}$, Eq.~(\ref{eq:fp0ci}).
A smaller value of $F_K/F_\pi$ corresponds to a smaller value of 
$L_5$ and thus of $\Delta(0)$. 
Note that varying $G'(0)$
within its error bar induces also a certain shift in the
results essentially for $\Delta_{CT}$ and $C_{12}$.

In order to illustrate the results, we reproduce in figure \ref{fig:RHC} the 
one shown 
in Ref. \cite{bops07} adding to the  
 dependence of $F_K /F_\pi$ and $f_+(0)$ on $\es- \ens$ and
$\delta+\ens$ the one of $\lambda_0$ using the central value of $\Delta (0)$. 
Note that while the errors
on $F_K /F_\pi$ and $f_+(0)$, which are purely 
experimental, are tiny, the ones on $\lambda_0$ 
coming from the two-loop ChPT calculations and not shown here are,
 as just discussed,
rather large. However, as can be seen from the figure,
a very precise knowledge of these three quantities would allow
to pin down the existence of physics beyond the SM.

As we have seen, the actual status of experiments and lattice results does not,
at present,
exclude the presence of physics beyond the SM in terms
of RHCs. As illustrated by the NA48 result, it could very well be
that $F_K/F_\pi$ and $f_+(0)$ is smaller, respectively larger than thought.
Interestingly this would lead to completely different values of the 
two $O(p^6)$ LECs $C_{12}$ and $C_{34}$.  
Since these enter other processes than the one discussed here  their study 
might help clarifying the situation. 
Clearly more work is needed on the lattice side as well as on the ChPT side to reach the needed accuracy.

\vspace{-0.3cm}
\section{Acknowledgements}

We are extremely grateful to  J. Stern for initiating the work and thank him 
for many useful comments
and discussions. The collaboration of M. Oertel at an early stage
is also acknowledged. We would like to thank Ulf-G. Mei{\ss}ner
for careful reading of the manuscript and G. Colangelo, J. Gasser, 
H. Leutwyler and C. Smith for discussions.

\end{document}